\documentclass[11pt]{article}

\usepackage{fullpage,latexsym,amsmath,amsfonts}

\def\01{\{0,1\}}

\begin{document}

\title{The Potential Impact of Quantum Computers on Society}
\author{Ronald de Wolf\thanks{QuSoft, CWI and University of Amsterdam, the Netherlands. Partially supported by ERC Consolidator Grant 615307 QPROGRESS.}}
\date{}
\maketitle

\begin{abstract}
This paper considers the potential impact that the nascent technology of quantum computing may have on society. It focuses on three areas: cryptography, optimization, and simulation of quantum systems. We will also discuss some ethical aspects of these developments, and ways to mitigate the risks.
\end{abstract}

\section{Introduction}

Quantum mechanics is our best physical theory for describing and predicting the behavior of small (and maybe also larger) objects. It was developed in the first quarter of the 20th century by people like Planck, Einstein, Bohr, Heisenberg, and Schr\"{o}dinger.
The predictions it makes about the behavior of small quantum systems are often weird and counter-intuitive, but have never been contradicted by experiments.
Computer science as a formal discipline started in the 1930s with the work of Church, Turing~\cite{turing:compnumb} and others.\footnote{The first digital computers were built a few years later. Of course, there was already much earlier (informal) work, like Pascal and Leibniz's 17th century work on machines for arithmetic calculations, Babbage and Lovelace's 19th century work on the Analytical Engine, and even the still-mysterious 1st century~BC Antikythera mechanism.} Its impact of ever-faster and ever-smaller computers on our society is already enormous, and still growing and deepening further.

The idea of using quantum mechanics to fundamentally improve the speed and/or security of computers has been around as a great idea on the crossroads of physics, computer science, and mathematics for nearly four decades. It has been a significant area for two decades, and it may start to become feasible technology in the next decade.  
Without going into the mathematics, we can identify three quantum-mechanical effects that make quantum computers tick:
\begin{itemize}
\item {\bf Superposition} allows a quantum computer's memory to be in many classical states ``at the same time,'' each state having a certain weight or ``amplitude'' (which may be negative). For example, a quantum bit (``qubit'') is a superposition of the two classical values 0 and~1. 
\item {\bf Interference} allows different superpositions to combine in a way that is similar to waves: positive and negative amplitudes will cancel each other out, while amplitudes with the same sign will add up.
\item {\bf Entanglement} allows different parts of the quantum computer, or even different quantum computers that are far away from each other, to be correlated in ways that are not possible classically.
\end{itemize}
These effects enable types of computation that are very different from---and in some cases much faster than---our ordinary, ``classical'' computers. Large quantum computers have not been built yet (the biggest ones at the time of writing have around 10--20 qubits), but a massive effort and investment is underway in academia and companies like Google and IBM to build them in the next decade or so.

Contrary to what one sometimes reads in popularizing accounts, quantum computers do not lead to exponentially faster computation across the board (nor does entanglement lead to faster-than-light communication). In fact, the set of computational problems where we know that quantum computers enable substantial speed-up is still quite small, and we can in fact prove that for many other problems quantum computers are no better than classical computers (see Aaronson~\cite{aaronson:limitsqc} for an accessible first introduction to the limits of quantum computers). One can think of a quantum computer as a massively complicated box, with exponentially many computations going on inside of it, whose combined outcomes we can only look at in very limited ways, like peeking inside the box through a few tiny holes.

Quantum computing, beyond its potential technological impact in the form of faster and/or more secure computers, already has an impact on fundamental physics (for example, quantum information theory and entanglement are gaining ground as an important viewpoint for material sciences, and even for understanding black holes~\cite{harlow&hayden:firewalls}), and on mathematics and computer science (as a new computing paradigm, but also as a new source of mathematical proof techniques~\cite{drucker&wolf:qproofs}). 
It also has a bearing on philosophical questions, from metaphysics (David Deutsch argues that a quantum computer would be a vindication of the many-worlds interpretation of quantum mechanics~\cite{deutsch:fabric}) to epistemology (our notions of ``knowledge'' and ``information'' should be adjusted to the weird proprieties of quantum superposition and entanglement) and ethics.

This paper will assume quantum computers of significant size will be realized at some point, and will try to chart the potential societal impact this will have. In other words, it will look at the areas where quantum computing may really have an impact beyond the realms of academia. In our view, the three main areas of potential impact are cryptography, optimization, and simulation of quantum systems, and we will examine these in the next section. After that we will look into ethical aspects of this impact, and ways to mitigate the risks.

\section{Potential societal impact}

\subsection{Breaking and remaking cryptography}

Cryptology is the art and science of secure communication in the presence of adversaries who try to eavesdrop on secret communication, forge identities, etc. Cryptology has a constructive branch (crypto\-graphy) and a destructive branch (cryptanalysis). The first branch designs schemes to ``encrypt'' information in such a way that only the intended parties can access it, leading to secure communication, e-commerce, unforgeable digital signatures, etc. The second branch tries to break, or ``decrypt'' such schemes. The productive tug of war between cryptographers and cryptanalysts (codemakers and codebreakers) has been going on for decades, some would even say for millennia: Suetonius~\cite[Divus Julius, \S 57]{suetonius:lives} already describes a simple scheme used by Julius Caesar to encrypt his letters: shift all characters in the alphabet by~3, so `a' becomes `d', `b' becomes `e' etc. This scheme is so weak that once one knows the shift (maybe after trying out all 25 possibilities) one can decrypt the texts. Since then, especially since the 1970s, much more sophisticated encryption schemes have been designed and deployed in practice.

Roughly speaking, much of cryptography is based on mathematical problems that are easy to compute in one direction, but hard to compute in the other direction.\footnote{In the theory of computing, a computational problem is considered ``easy'' if it can be computed by an algorithm whose running time grows at most polynomially with the input length (i.e., a running time like $n^2$ or $n^3$ for inputs of $n$ bits). Otherwise the problem is considered ``hard''; often such hard problems take a running time that grows exponentially or nearly-exponentially with the input length~$n$. The latter type of problem is not solvable in reasonable amounts of time for large input length.} A prime example of this (no pun intended) is multiplication: it is very easy to multiply together two large numbers $p$ and $q$ to obtain their product $N=p\cdot q$; but going the other way, computing the factors $p$ and $q$ from $N$, is believed to be a hard computational problem for \emph{classical} computers. How can we use easy-one-way/hard-the-other-way problems for cryptography? The basic idea of the RSA crypto-system is as follows. Suppose Alice wants to enable the rest of the world to send her encrypted messages, that only she can decrypt. She can choose two large prime numbers $p$ and $q$ (her \emph{secret} key), and only make public their product $N$ and an associated ``encryption exponent'' (this is her \emph{public} key). Using the public key, anybody else can encrypt their messages to Alice in such a way that Alice can decrypt the messages using the extra information of her secret key. In contrast, classical eavesdroppers without this extra information can learn the encrypted message by tapping the communication channel, but cannot (as far as we know) decrypt the message in any reasonable amount of time. 

The first big hit of quantum computers was Peter Shor's 1994 efficient quantum algorithm for finding the prime factors of large numbers~\cite{shor:factoring}.  
A sufficiently large quantum computer would thus be able to compute the secret key from the public key of the RSA scheme, and hence can decrypt the encrypted messages sent to Alice. While some parts of classical cryptography are not affected by such attacks, much of our online communication, e-commerce etc., is protected by cryptographic schemes based on the hardness of factoring or similar problems that can also be efficiently solved by a quantum computer, such as ``discrete logarithms.'' 
Shor's discovery was really the point when the area of quantum computing started to move from a fringe activity to a central area of physics and computer science, with lavish attention from funding agencies (interested in both basic science and technology) as well as spy agencies (interested in breaking enemy codes).

The breaking down of much of our cryptography would have a large impact on our economy and society, much of which assumes we can safely communicate, transfer money, sign documents electronically, etc.  A world without reliable electronic payments and bank transactions would come to a grinding halt---clearly cash payments or barter are not good alternatives. Governments and many other organizations rely crucially on the ability to communicate secretly. Even if it will take decades to actually build a quantum computer big enough to factor large numbers, for things that have to remain secret for the next 20 to 30 years (a typical requirement for government secrets), the future quantum threat is already an acute problem now: enemy spies  or the mafia can already hoover up encrypted communication today, store it, and decrypt it later when a quantum computer becomes available. 

We mention two ways to remedy this, neither of them ideal.
The first is so-called \emph{post-quantum cryptography}. This is classical cryptography, based on computational problems that are easy to compute in one direction but hard to compute in the other direction \emph{even by quantum computers}. Factoring does not fit this bill because of Shor's quantum algorithm, but there have been proposals for using other computational problems, for instance based on lattices or on error-correcting codes. The problem with such schemes is that they have barely been tested. We are not able to \emph{prove} that factoring is a hard problem for classical computers, but at least one good piece of evidence for such computational hardness is that many sharp mathematicians have tried for decades to find efficient factoring algorithms, and failed. The alternative computational problems that have been suggested for post-quantum cryptography, have not yet undergone such scrutiny and there may well exist an efficient quantum (or even classical!) algorithm for breaking them. 

The second way to remedy the quantum attack on much of current-day cryptography is to use \emph{quantum cryptography}, which uses quantum effects to design more secure cryptographic systems. The key property is the fact that measuring an unknown quantum state will disturb it, and such disturbance can be detected by the honest parties. The most famous example of such quantum cryptography is the BB84 ``quantum key distribution'' scheme of Bennett and Brassard~\cite{bb84}: using quantum communication, Alice and Bob (who trust each other but not the quantum channel over which they communicate) can either establish a shared secret key unknown to any eavesdropper, which they can then use for secure communication---or they can detect the presence of the eavesdropper. This scheme can be proved\footnote{Under several idealizing assumptions that are approximately true in practice: quantum mechanics is the correct description of Nature; Alice's lab and Bob's lab is secure from the eavesdropper; their communication channel is ``authenticated'' (they know they're talking to one another); and their apparatuses have low and benign errors.} secure, even against quantum adversaries with unlimited amounts of time and computing power. 
Because the required hardware for such quantum cryptography is much simpler than for a large quantum computer, commercial implementations of these schemes already exist. These have repeatedly been hacked due to practical imperfections in their implementation, but they are getting better.% 
\footnote{If neither post-quantum nor quantum cryptography works, then as a last resort one can always put one's secrets on a high-quality memory device detached from the internet and put this (or even a print-out) in a physical safe. However, this has many obvious disadvantages over computer-based cryptography.}

\subsection{Faster search and optimization}

Governments, companies, and other organizations often use their computers to solve large search or optimization problems: to finding efficient allocations of resources to the tasks they need to solve, to schedule work (or classes, like in schools and universities), to search through large data files, to design energy-efficient chips or airplanes, etc.
For many such tasks, quantum computers can offer significant speed-ups, 
for example for search problems~\cite{grover:search}, finding the minimum or maximum of a given function over some finite domain~\cite{durr&hoyer:minimum}, finding the shortest route between two points on a map~\cite{dhhm:graphproblemsj}, approximately solving large systems of linear equations (in a somewhat weak sense)~\cite{hhl:lineq}, as well as other problems. See~\cite{montanaro:algosurvey} for a recent survey.
%, backtracking algorithms~\cite{montanaro:backtracking}, linear and semidefinite programming~\cite{brandao&svore:qsdp}.

The area of \emph{machine learning} deserves special mention here. This area has really taken off in the last 5 years or so, with spectacular progress in areas like image recognition, natural language processing, and even in beating the best human players at games like Go and poker. This is all based on classical computers.
Usually in machine learning one is given data and one has to find some model or hypothesis that ``fits'' this data well. Finding a well-fitting or even optimally-fitting model for the data is an optimization problem that can sometimes be solved better or faster on a quantum computer using the techniques mentioned above. 

One problem with most of these approaches is that they typically need to be able to access, in quantum superposition, large amounts of \emph{classical} data. Having very large quantum-addressable classical memories is technologically very demanding, and will not be realized any time soon.
Also, the speed-up these methods give over the best classical approaches is usually only ``polynomial'' (typically quadratic or less); it is not the exponential improvement that Shor's algorithm gives over the best known classical algorithms for factoring large numbers. Still, even such polynomial speed-ups can make a big difference in practice particularly when applied to very large inputs.

\subsection{Simulating quantum systems}

A third area where quantum computers are likely to have an impact, is in simulating the behavior of quantum systems. This was in fact the main reason for Richard Feynman to dream up the idea of quantum computing in the first place~\cite{feynman:simulating}: a quantum state involving $n$ particles has a number of parameters that is exponential in $n$, so it seems plausible that simulating its behavior on a classical computer takes exponential time (though, like many plausible things in complexity theory, we do not know how to prove this). In contrast, a quantum computer can in principle simulate the behavior of any other quantum system efficiently: one would effectively write a program with the physical laws acting on the quantum system that one wants to simulate, and then let the quantum computer execute this program to see what happens. This way, a quantum computer with, say, electron-based qubits could mimic the behavior of other quantum systems, such as specific atoms or molecules.

A big chunk of the computing time of supercomputers today is spent on simulating quantum systems, and a quantum computer could make a huge difference here.
Some applications, such as figuring out properties of specific molecules that are beyond the reach of our current best computers, could already be addressed by a quantum computer with a few hundred well-behaved qubits~\cite{poulinea:trottersize,wla:elecstruc}.\footnote{In contrast, running Shor's algorithm to break current cryptography would require thousands or even millions of qubits, depending on how error-free we can make these qubits and the operations upon them.}

Further applications occur where we do not just want to simulate the behavior of one specific quantum system, but are looking for a quantum system with certain desirable properties.
A good example is drug design: when pharmaceutical companies look for drugs against specific afflictions, they basically look through enormous lists of possible molecules, trying each one out (in simulation or experiment) to find one that has the right properties. 
Quantum computers can help in two interlocking ways: they can figure out the properties of a specific molecule faster 
than classical computers can, and they can search through large sets of promising molecules for one that actually has the desired behavior, using Grover's search algorithm or some other quantum method for optimization.
Also the design of more efficient materials and more efficient production of fertilizer~\cite{reiherea:fertilizer} are often mentioned as potential applications here. Fertilizer is so important for the world's food production that even a small efficiency improvement would already be very valuable.

\section{Ethical aspects}

\subsection{Cryptography}

Like any technology, cryptography can be used for good or bad: to protect activists against an overbearing government, but also by terrorists for planning attacks without being overheard. The same is true for cryptanalysis: it can be used by the NSA to thwart those terrorists, but also to conduct mass-surveillance and read the emails of millions of innocent people.
Clearly, a breakdown of cryptography due to quantum computers will have serious implications. Much information that is now kept secret for good reason would be out on the street. Privacy would be much diminished. Governments would have great trouble protecting their workings against foreign spies.
The right balance between privacy and justified surveillance for security purposes is a very tricky ethical question, and this balance changes if the set of available cryptographic tools changes. Quantum cryptography may become available, but is likely to be both much more expensive and more clunky than RSA, and may only be within reach of governments and large organizations.

\subsection{Increased inequality}

What about the other two areas of potential impact, faster search/optimization, and simulation of quantum systems? From a utilitarian perspective, increased efficiency in areas like planning, resource allocation, machine learning, and development of new drugs and materials, seems mostly a good thing: reducing waste, freeing up time, increasing profit, curing diseases, etc. Of course, such efficiency gains could also facilitate terrorism, mass-surveillance, and other undesirables, but this caveat applies to better technology in general. Indeed, if these efficiency gains due to quantum computers are going to be widespread and accessible to all, there would be little to say about their specific ethical aspects.

It is anybody's guess how quantum computing power will spread through society in the next few decades, but it is quite possible that it will \emph{not} be widely accessible. This could lead to a more unequal distribution of power and wealth: between America and the rest of the world, and between a few big companies and the rest of society. 

First, most of the companies that have started to invest heavily in quantum computing (hardware as well as software) are American: IBM, Google, Microsoft, and Intel. These companies are already patenting many of the ideas, including some based on freely-available academic research. This could lead to American businesses dominating (or even monopolizing) commercial quantum computing, similar to the way Silicon Valley dominates much of current classical computing. 
Even worse, because quantum computing development takes massive investment, entrenched giant companies are less likely to be disrupted by competition from small start-ups than in classical computing.
The European Union is trying to counter this potential American dominance by promoting collaboration between European academia and European companies as part of its recent Quantum Flagship initiative, but Europe does not have tech companies of quite the same heft as the US giants.
%, nor quite the same risk-taking start-up culture. 
Moreover, at the level of government agencies, few countries (except maybe China and the EU as a whole) can match the secret but massive budgets available at the NSA, some of which are used for quantum computing research. 
 
Second, even if American companies and/or government do not end up dominating the development of the hardware and software of quantum computing, wide accessibility of quantum computing power is not guaranteed. Certainly initially it will be exceedingly expensive to build even one medium-size quantum computer. The risk here is that only a few large companies will be able to afford quantum computers, will use the efficiency gains to outcompete their competitors, and form monopolies or oligopolies. For example, if quantum computers turn out to be great at developing new drugs, and only one pharmaceutical company has access to a quantum computer, it could dominate its industry. This would weaken competition in our free-market economy, shift power from individuals and small businesses to one or a few large companies, and lead to growing inequality in society.

\subsection{Making the impact positive}

The same above-mentioned risks of increased inequality were present in the 1950s and 1960s, when few companies were able to afford the new mainframe computers peddled by the likes of IBM. Since then, of course, the wonders of Moore's law have made classical computing power available to most people---first through PCs in the 1980s, and more recently in the form of cheap smartphones that have more computing power than even the strongest machines from the 1960s. Interestingly, nowadays one can even cheaply ``rent'' massive computing power through the cloud from Amazon and other companies.

The hope is that the future of quantum computing goes along the same lines: first a few very expensive and moderately powerful machines, and then subsequent generations of cheaper and more powerful quantum computers. It is likely that quantum computers will remain rather bulky and expensive in comparison with today's computers, and it is neither likely nor necessary that many people will have one on their desk. The best way to make quantum computing power widely available is to build a few very strong quantum computers, and to enable people to run programs on them through the cloud, maybe for a fee. IBM's ``Quantum Experience''~\cite{IBM:qexp}, which lets everyone play around with a 5-qubit computer through the cloud for free, is a toy version of this idea. Wide accessibility of quantum computing power might even lead to a new twist on the ``maker movement,'' where lots of enthusiastic amateurs start to simulate and play around designing new molecules, new materials etc., thereby actually reducing the power of big pharmaceutical and other companies as compared to today.

Ideally such accessibility would come about for commercial reasons, much like Amazon's cloud services. If this does not happen and companies start hoarding quantum computing power, other parties (governments or otherwise) should try to rectify this. Unfortunately, governments themselves could also have an interest in monopolizing access to quantum computers. However, one large quantum computer that is made accessible through the cloud by some organization for commercial or benevolent reasons (say by the Gates Foundation, or by a country like Norway) is already enough to counter many of the risks of monopolization of access to quantum computing. 

In addition to access to the computing power, there is also the issue of access to \emph{knowledge}. Until recently, most of the knowledge produced about quantum computing (with the exception of some research by intelligence agencies like the NSA) was made publicly available in the form of academic research papers. More recently, with the growing influence of large companies, there is a bigger push to patent knowledge or even to hoard it. Researchers both in academia and industry have a responsibility to counteract this by insisting on publication of their work, maybe after a short delay for commercial reasons.
This is justified by the general reason why science should be open: it leads to faster spreading and development of knowledge, as well as to faster detection and correction of errors.

\section{Summary and conclusion}

Quantum computers are getting closer to realization, and they could have significant impact on society in several areas. First, they will be able to break much of current cryptography, endangering our digital economy, though they also provide cryptographic alternatives. 
Second, they will be able to optimize all sorts of processes better, leading to efficiency gains.
Third, they will allow for much faster simulation of quantum-mechanical systems, with the potential of better design of drugs, materials, etc.

We also briefly considered the ethical aspects of this. 
In destroying much of classical crypto\-graphy, quantum computers
reduce online privacy and make it harder to hide things (both for good and for bad purposes).
If access to quantum computers is limited to one or few governments, it could upset the balance of power between different countries;
and if this access is limited to a few big companies it could lead to monopolies or oligopolies, increasing inequality in society.\footnote{These ethical issues are not really specific to \emph{quantum} computers---they would apply also in case of much faster \emph{classical} computers and/or much better \emph{classical} algorithms. However, we expect the improvements in classical hardware to slow down (the end of Moore's law is near), and we do not expect much faster classical algorithms for problems like factoring large numbers or simulating quantum systems.}

We feel the best way to ensure that the positive impact of quantum computers outweighs the negative is through openness and accessibility: scientific knowledge should be made publicly available as much as possible, and quantum computing power should be made accessible through the cloud. Both scientists and governments (and companies as well, partially through the laws issued by those governments) have a responsibility here.

\paragraph{Acknowledgment:}
Thanks to Joran van Apeldoorn, Harry Buhrman, and the anonymous referees for helpful comments that improved the presentation of this paper.

\bibliographystyle{alpha}
%\bibliography{qc}

\newcommand{\etalchar}[1]{$^{#1}$}

\end{document}